\documentclass[ms,11pt]{egu}       
\usepackage{amsmath}
\usepackage{color}

\usepackage{graphicx}

\printfigures              

\begin{document}

\title{Scaling collapse and structure functions:
 Identifying self-affinity in finite length time series.}

\author[1]{S. C. Chapman, B. Hnat, G. Rowlands}
\author[2]{N. W. Watkins}

\affil[1]{Space and Astrophysics, University of Warwick, UK}
\affil[2]{British Antarctic Survey (NERC), Cambridge, UK}

\runningtitle{Scaling collapse} \runningauthor{Chapman et. al.}

\correspondence{S. C. Chapman (sandrac@astro.warwick.ac.uk)}

\journal{\NPG}       

\date{May 2005}


\msnumber{npg-2005-0014}

\maketitle

\begin{abstract}
Empirical determination of the scaling properties and exponents of
time series presents a formidable challenge in testing, and
developing, a theoretical understanding of turbulence and other
out-of-equilibrium phenomena. We discuss the special case of self
affine time series in the context of a stochastic process. We
highlight two complementary approaches to the differenced variable
of the data: i) attempting a scaling collapse of
  the Probability Density Functions which
should then be well described by the solution of the corresponding
Fokker-Planck equation and ii) using structure functions to
determine the scaling properties of the higher order moments. We
consider a method of conditioning that recovers the underlying
self affine scaling in a finite length time series, and illustrate
it using a L\'{e}vy flight.
\end{abstract}

\introduction\label{sec:intro}

 Theories of turbulence as applied to single point
measurements in a flow concern the scaling
  properties, in a statistical sense, of differenced time series, where the Taylor hypothesis is
   invoked so that the difference between measurements at some time $t$ and
   a later time $t+\tau$ acts as a proxy for the difference between measurements made
    at two points in the fluid separated by length scale $L$.
Studies of scaling in solar wind turbulence have focused on the
power spectra and the structure functions (see e.g.
\cite{tumarsh,horbury}) and, more recently, the Probability
Density Function (PDF) (\cite{hnatBgrl,hnatpre}).

The statistical scaling properties of time series can in general,
however,  be
    considered in a similar manner.
There is  a considerable literature concerning  scaling in auroral
region magnetometers and in geomagnetic indices (such as
\cite{tsurutani,takalo93,consolini96,voros98,uritsky98,watkins,kovacs}).
This is motivated in part by attempts to understand the driven
magnetospheric system from the perspective of scaling due to
intrinsic processes (see e.g. \cite{chapman01} and references
therein) and their relationship to that of the turbulent solar
wind driver. This necessitates quantitative \textit{comparative}
studies of scaling in time series (e.g.
\cite{takalo98,freeman,uritsky,voros02,hnatgrl}). Such studies can
to some extent fruitfully consider the low order moments, whereas
a particular difficulty for comparison of observations with models
of turbulence is that the intermittency parameter in turbulence
$\tau(2)$ is determined by the $6^{th}$order structure function
\cite{frisch}.

More recently, studies have focussed on the scaling properties and
functional form of the PDFs of the differenced time series (see
e.g. \cite{consolini98,valvo,weigel03a,stepanova}). This leads to
a Fokker-Planck model in the case of  self- similarity
(\cite{hnatpre,hnatjgr}).

In this paper we describe an approach to modelling such scaling
data which exploits the data's self-affine property by applying
the idea of coarse graining the data (\cite{wilson,sornette}),
here, in the time domain. This coarse-graining can be achieved
empirically, from the data, by a scaling collapse procedure
(e.g.\cite{hnatpre,hnatjgr}) (section 2), and, then having
experimentally determined the scaling exponent, we can take the
approach one stage further and seek to describe the data by means
of a particular case of a generalised Fokker-Planck equation
(GFPE, section 5). We stress here that the GFPE is here, as
elsewhere (eg \cite{sornette}), applied to a much more general
class of problem than the strictly equilibrium physics for which
the original FPE was obtained. The GFPE represents an alternative
to the fractional Fokker-Planck equation (e.g. \cite{zav}) which
is also applicable in such non-equilibrium cases.

The critical steps in this process are then (i) establishing
whether a given dataset is self affine and (ii) determining the
scaling exponent. We highlight two important issues that arise in
the analysis of physical datasets here.

The first of these is that SDE models for the data, and indeed,
coarse graining, deal with the properties of an arbitrarily large
dataset. We use a well understood example of a self affine
timeseries, that of ordinary Levy motion (section 3), to show how
conditioning of the data is needed to recover the known scaling of
an arbitrarily large timeseries from one finite length. We then
use an example of a naturally occurring timeseries, that of the AE
geomagnetic index, shown previously to exhibit self affine scaling
over a range of timescales, to highlight the effectiveness, and
the limitations, of this technique.

The second of these is that knowledge of the scaling properties of
(in principle all) the non zero moments is needed to capture the
scaling properties of a timeseries. We again use the AE timeseries
to illustrate this point by constructing a fractional Brownian
motion fBm with the same second moment, but with a very different
PDF.

\section{Self affine time series: concepts} \label{renorm}

   From a time series
$x(t)$ sampled at times $t_k$, that is at evenly spaced intervals
$\Delta =t_k-t_{k-1}$ we can construct a differenced time series
with respect to the time increment $\tau=s\Delta$:
\begin{equation}
  y(t,\tau ) = x(t + \tau ) - x(t)
 \end{equation}
 so that
 \begin{equation}
 x(t + \tau ) = x(t) + y(t,\tau)\label{walk}
 \end{equation}
If we consider $N$ successive values determined at
 intervals of $\Delta$, that is,
 $y(t_1,\Delta)..y(t_k,\Delta)...y(t_N,\Delta)$,
  their sum gives:
\begin{equation}
x(t)=\sum_1^N y(t_k,\Delta)+x_0\label{ysum}
\end{equation}
where $x_0=x(t-N\Delta)$.  As $N\rightarrow \infty$
 the sum (\ref{ysum}) of the $y$ tends to the original time series $x(t)$.

We  will make two assumptions: i) that the $y(t,\tau )$ is a
stochastic variable so that (\ref{walk}) can be read as a random
walk and ii) that the $y$  are scaling with $\tau$ (to be defined
next).

  By summing adjacent
 pairs in the sequence, for example:
 \begin{equation}
y^{(1)}(t_1,2\Delta)=y(t_1 ,\Delta ) + y(t_2 ,\Delta )
\end{equation}
one can coarsegrain (or decimate) the time series in $\tau$. This
operation gives the $x(t)$ as a random walk
 of $N/2$ values of $y$
 determined at intervals of $\tau=2\Delta$. We can successively
 coarsegrain the sequence an arbitrary number of times:
\begin{eqnarray}
x(t)&
 = &y(t_1 ,\Delta ) + y(t_2 ,\Delta ) + \cdots + y(t_k ,\Delta ) + y(t_{k + 1} ,\Delta ) + \cdots + y(t_N ,\Delta )
 \\\nonumber
 & =& y^{(1)} (t_1 ,2\Delta ) + \cdots + y^{(1)} (t_k ,2\Delta ) + \cdots + y^{(1)} (t_{N/2} ,2\Delta )
 \\\label{rg}
 & = &y^{(n)} (t_1 ,2^n \Delta ) + \cdots + y^{(n)} (t_k^{} ,2^n \Delta ) + \cdots + y^{(n)}
  (t_{N/2^n } ,2^n \Delta )\nonumber
 \end{eqnarray}
 where this procedure is understood in the renormalization sense
 in
 that both $N$ and $n$ can be taken arbitrarily large, so that a
 timeseries of arbitrarily large length is considered. This
 procedure can apply to a finite sized physical system of interest provided that that system
 supports a large range of spatio- temporal scales (the smallest
 being $\Delta$, the largest, $2^n \Delta$, $n$ large), an example
 of this is the inertial range in fluid turbulence.

We now consider a \emph{self affine} scaling with  exponent
$\alpha$:
\begin{equation}
 y' = 2^\alpha  y,\;\;\;\tau ' = 2\tau ,
 \end{equation}
 so that
 \begin{equation}
 y^{(n)}  = 2^{n\alpha } y, \;\;\;\tau  = 2^n \Delta
 \end{equation}
 For arbitrary $\tau$ we can normalize ($\tau\equiv\tau/\Delta$)
 and write
 \begin{equation}
 y'(t,\tau ) = \tau ^\alpha  y(t,\Delta )\label{scale}
 \end{equation}
 Now if the $y$ is a stochastic variable with self affine scaling
 in $\tau$, there exists a \emph{self similar}
  PDF which is unchanged under the transformation (\ref{scale}):
 \begin{equation}
 P(y'\tau ^{ - \alpha } )\tau ^{ - \alpha }  = P(y)\label{rescale}
 \end{equation}
 Importantly, the $y's$ are not necessarily Gaussian distributed  stochastic
 variables, but do possess self similarity as embodied by
 (\ref{rescale}).

 This property is shared by the ($\alpha-$stable) L\'{e}vy
 flights (\cite{slez}) for $N\rightarrow\infty$. The
special case where the $y's$ are both independent, identically
distributed (iid) and have finite variance corresponds to a
 Brownian random walk. One can show directly from the above renormalization
 (see for example \cite{sornette}) that the Brownian case is just the
 Central Limit Theorem with $\alpha=1/2$ and Gaussian $P(y)$.
 Here, we  consider time series which  possess the properties
 (\ref{scale}) and (\ref{rescale}), which may have $\alpha \neq 1/2$ and which are time stationary
 solutions of a Fokker-Planck equation.

An important corollary of (\ref{rescale}) is  of the scaling
of the structure functions (and moments). The $p^{th}$ moment can
be written as:
\begin{equation}
m_p=<y^p>=\int_{-\infty}^{\infty}P(y)y^pdy=\tau^{p\alpha}\int_{-\infty}^{\infty}P(y')y'^pdy'
\end{equation}
so that
\begin{equation}
m_p\sim \tau^{p\alpha}\label{momscale}
\end{equation}
via (\ref{rescale}).The scaling of any of the non zero moments of
a self affine time series is thus sufficient to determine the
exponent. Importantly, all the non zero moments will share this
same scaling. This can also be appreciated directly by writing the
PDF as an expansion in the moments. If we define the Fourier
transform of the PDF $P(z)$ of a given time series $z(t)$ by:
\begin{equation}
\hat{P}(k)=\int_{-\infty}^{\infty}e^{ikz}P(z)dz
\end{equation}
then it is readily shown that the $p^{th}$ moment is given by:
\begin{equation}
m_p=(-i)^p\frac{d^p\hat{P}(k)}{dk^p}\mid_{k=0}
\end{equation}
where $d^p/dk^p$ denotes the $p^{th}$ derivative with respect to
$k$. From this it follows that the PDF can be expressed as an
expansion in the moments:
\begin{equation}
\hat{P}(k)=\sum_{p=0}^{\infty} \frac{m_p}{p!}(ik)^p\label{momexp}
\end{equation}
Hence the PDF is defined by knowledge of \emph{all} the non zero
moments.

\section{Testing for self affine scaling.}
\subsection{Extracting the scaling of a surrogate, a
finite length L\'{e}vy flight.}

We now discuss methods for testing for the property
(\ref{rescale}) and measuring the  exponent $\alpha$ for a given
finite length time series. For the purpose of illustration we
consider a L\'{e}vy flight of index $\mu=1.8$ which is generated
from iid random deviates by the following algorithm for the
increments (the $y's$, see \cite{sieg} for details):
\begin{equation}
f_{\mu}=
\frac{\sin(\mu r)}{(\cos(r))^\frac{1}{\mu}}
\left(\frac{\cos[(1-\mu)r]}{v}\right)^\frac{(1-\mu)}{\mu}\label{levy}
\end{equation}
where $r$ is a uniformly distributed random variable in the range
$[-\pi/2, \pi/2]$ and $v$ is an exponentially distributed random
variable with mean $1$ which is independent of $r$. The scaling
exponent $\alpha$ from (\ref{scale}) and (\ref{rescale}) is then
related to the L\'{e}vy index, $\mu$, by $\alpha=1/\mu$.

One can first consider directly attempting a scaling collapse  in
the sense of  (\ref{rescale}), of the PDF of differences obtained
over a wide range of $\tau$  (see \cite{stanley,hnatgrl,hnatpre}
for examples). This corresponds to a renormalization of the data
as discussed above. We first determine the scaling exponent
$\alpha$ from one or more of the moments via (\ref{momscale}) or
an estimate thereof. In a finite length time series, one would
ideally use the scaling of the peak $P(y=0, \tau)$ (that is, the
$p=-1$ moment) with $\tau$ as this is better resolved
statistically than the higher order moments. In practice however
the time series $y(t,\tau)$, formed from the differences of a
measured quantity, can as $y\rightarrow 0$ be dominated by
observational uncertainties.

Figure 1 shows the scaling collapse (\ref{rescale}) applied to a
numerically generated L\'{e}vy flight (\ref{levy}) of $10^6$
increments. The curves correspond to differences at values of
$\tau=m \Delta$ with $m=[6, 10, 16, 26, 42]$. Error bars denote an
estimate of  the expected fluctuation per bin of this histogram
based on Gaussian statistics (a more sophisticated method for
estimating these for the L\'{e}vy case may be found in
\cite{sieg}). We see that scaling collapse can be verified to the
precision with which the PDF is locally determined statistically.
\begin{figure}[t]
\vspace*{2mm}
  \centering{\includegraphics[width=8.3cm]{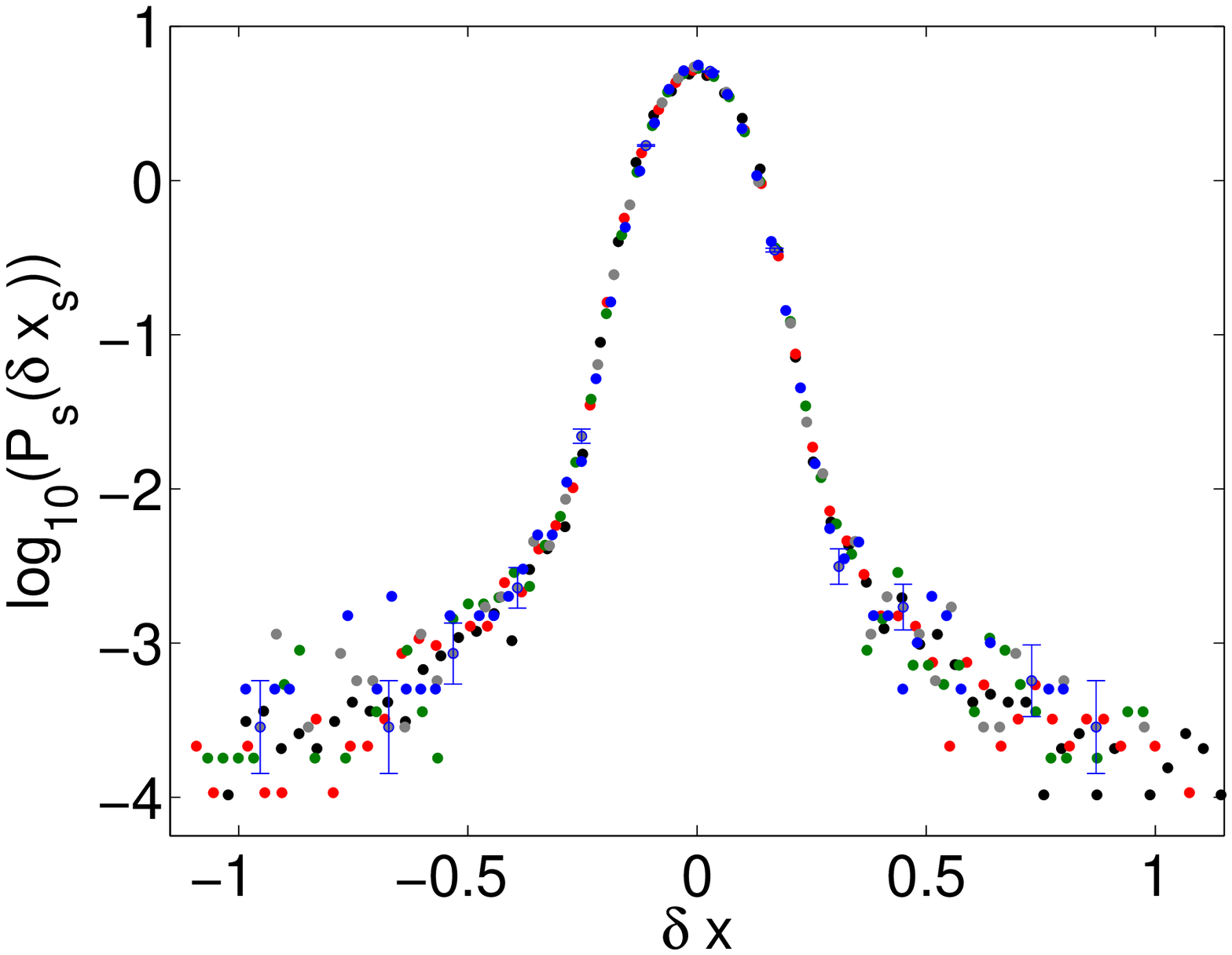}} 
  \caption{Scaling collapse of the PDF of an $\mu=1.8$ L\'{e}vy
  flight.
    \label{fig:1}
    }
\end{figure}
The exponent $\alpha=0.544$ used to achieve the scaling collapse
in Figure 1  was determined empirically directly from an analysis
of this finite length time series based on the structure functions
 discussed below.

As discussed above, the scaling exponent $\alpha$ that
successfully collapses the PDF of different $\tau$ should emerge
from the scaling of the moments. This is often obtained via the
 generalized structure
 functions (see e.g. \cite{tumarsh,horbury,hnatgrl,hnatjgr} for examples)
\begin{equation}
 S_p (\tau ) =  < |y(t,\tau )|^p  >  \propto \tau ^{\zeta
 (p)}\label{sfns}
 \end{equation}
 where for  self affine $y(t)$, we have $\zeta (p)= p\alpha$ (for a multifractal, $\zeta (p)$ is approximately quadratic in $p$).
From (\ref{momscale}) the moments will in principle share this
scaling provided that the moment is non- zero (however in a noisy
signals a moment that should vanish will be dominated by the
noise). In principle we can  obtain $\alpha$ from the slopes of
log- log plots of the $S_p$ versus $\tau$ for any $p$; in practice
this is severely limited by the finite length of the dataset.

\begin{figure}[t]
\vspace*{2mm}
  \centering{\includegraphics[width=8.3cm]{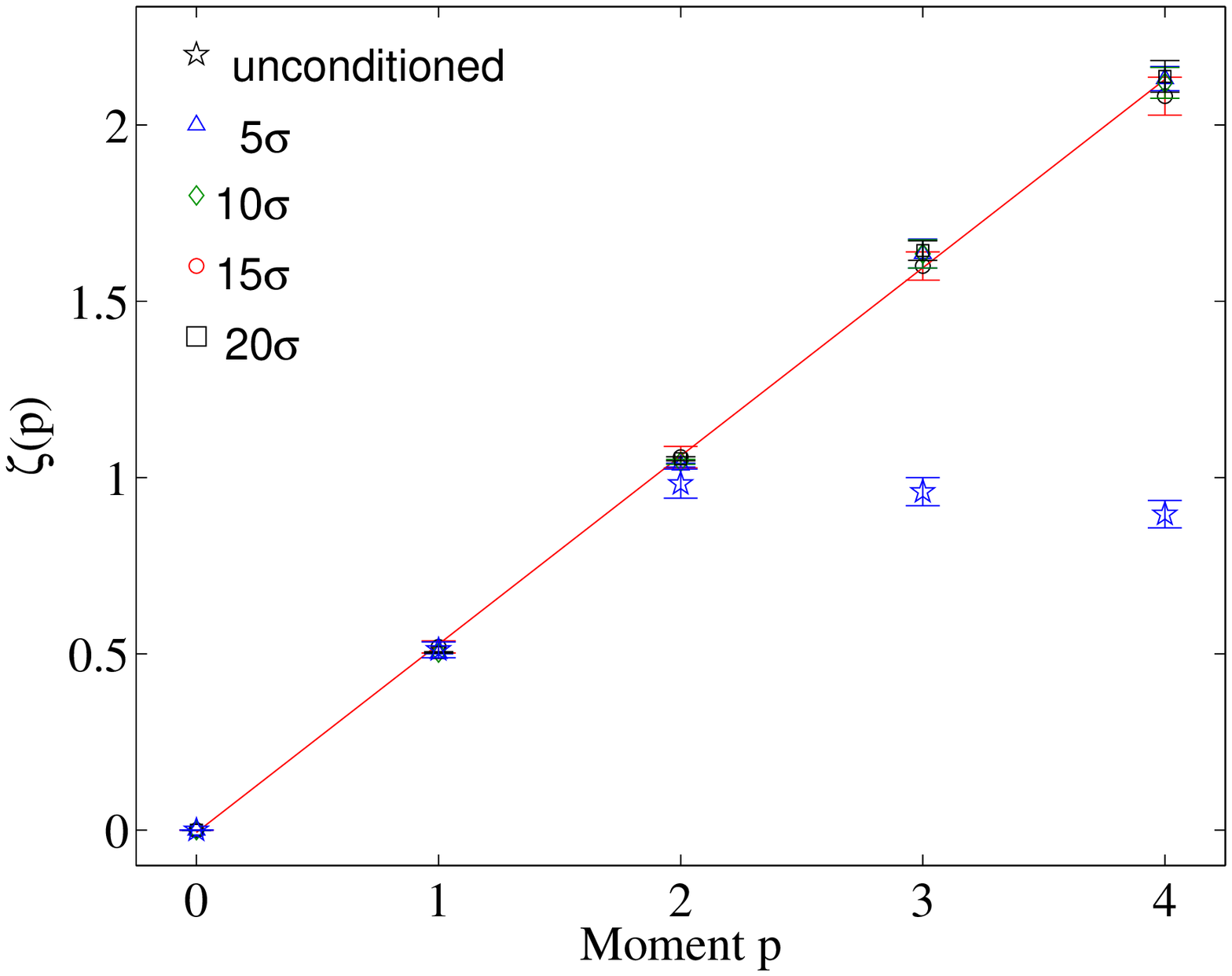}} 
  \caption{The effect of conditioning a Levy flight. $\bigstar$
  are the $\zeta(p)$ obtained from the raw time series, all other
  symbols refer to conditioned time series for different values of $Q$ (see text). The
  conditioned results yield a scaling  exponent $\alpha =0.544$
  which corresponds to a Levy index of $\mu =1.84$.
    \label{fig:2}
    }
\end{figure}
The $\zeta(p)$ for the above L\'{e}vy flight obtained via
(\ref{sfns}) are shown plotted versus $p$ in Figure 2. On such a
plot we would expect a straight line $\zeta(p) \sim p\alpha$ but
we see here the well known result (see for example
\cite{levy1,levy2}) that for the surrogate, the L\'{e}vy time
series of finite length, there is a turnover in scaling above
$p=2$ which is spurious in the sense that it does not reflect the
self affine scaling of the infinite length timeseries.

One way to understand   this spurious  bifractal scaling is that
in a finite length time series the PDF does not have sufficient
statistical resolution in the tails. Infrequently occurring large
events in the tails will tend to dominate the higher order
moments. We need to eliminate those large events that are poorly
represented statistically without distorting the scaling
properties of the time series. For a self affine time series an
estimate of the structure functions is:
 \begin{equation}
  S^C_p= \int_{ - A}^A |y|^p P(y,\tau )dy \approx<  |y|^p  > \label{cond}
\end{equation}
where the limit on the integral is proportional to the standard
deviation $\sigma$ so that $A = Q\sigma (\tau )$, with some $Q$
constant. Now $\sigma(\tau)\sim\sqrt<y^2>\sim \tau^\alpha$ shares
the same self affine scaling with $\tau$ as the original
timeseries $y(t,\tau)$, so that if $S_p\sim \tau^{p\alpha}$ under
(\ref{rescale}) then, importantly,  $S^C_p\sim \tau^{p\alpha}$
also. Provided that $Q$ can be chosen sufficiently large to
capture the dynamic range of $y$, and provided that $P(y)$ is
symmetric, (\ref{cond}) will provide a good estimate of $\alpha$.
This is demonstrated in figure 2 where we also show the $\zeta(p)$
obtained from (\ref{cond}).

One can thus see that once a conditioning threshold is applied,
the self affine scaling of the L\'{e}vy flight is recovered and
the value of the scaling exponent
  is insensitive to the value of $Q$ chosen (for $Q$
sufficiently large). We obtain the value of $\alpha=0.544$ used
for the scaling collapse in Figure 1 once conditioning is applied,
giving an estimate of $\mu=1.84$, consistent with the index used
to generate the synthetic L\'{e}vy flight (\ref{levy}). Similar
results for a surrogate Levy dataset have been obtained by M.
Parkinson (private communication, 2004).

An analogous procedure to (\ref{cond}) can also be realized by
means of a truncated wavelet expansion of the data (see for
example \cite{kovacs,salem}).

 In (\ref{cond}) we assumed self affine
scaling in choosing the functional form of the limits of the
integral. In a given time series the scaling may not be known a
priori. If for example the time series were multifractal
($\zeta(p)$ quadratic in $p$)
 we would obtain from
(\ref{cond}) a $\zeta(p)$ which varied systematically with $Q$. In
practice, several other factors may also be present in a time series
which may additionally reduce the accuracy of the approximation
(\ref{cond}).

\subsection{Extracting the scaling of a 'natural' example, the AE
timeseries}

To illustrate the above, we consider an interval of the AE index
shown previously to exhibit weakly multifractal scaling
(\cite{hnatjgr}). The scaling index is not within the L\'{e}vy
range and thus it has been modelled with a GFPE rather than a
L\'{e}vy walk \cite{hnatjgr}.

The PDF of differenced AE is asymmetric \cite{hnatgrl}, and the
scaling in $\tau$ is broken as we approach the characteristic
substorm timescale of 1-2 hours. Remnants of the substorm
signature will be present in the time series on timescales shorter
than this. The behaviour of the peak of the PDF ($P(y\rightarrow
0)$) will also be dominated by uncertainties in the determination
of the signal rather than its scaling properties.

Figure 3 shows a plot of $\zeta(p)$ versus $p$ for the AE time
series in the same format as figure 2 for the interval January
1978 to July 1979 comprising $7.5 \times 10^5$ samples. Plots of
the structure functions used to construct figure 3 are shown in
figure 4. The error bars on figure 3 are those of the best fit
straight lines to Figure 4 rather than the possible range of
straight line fits and as such are a minimal error estimate.

We plot in figure 4(a) the raw result, that is (\ref{sfns}) and in
figure 4(b) the conditioned approximation (\ref{cond}) with
$Q=20$, the latter corresponding to the removal of less than 1 \%
of the data. From figure 4 we see that no clear scaling emerges
beyond the third order $p=3$  until approximation (\ref{cond}) is
made. Clearly, if scaling is present, the $\zeta(p)$ obtained from
the raw structure functions (figure 4(a)) are not a good estimate.
Once the data is conditioned, we find that $Q=[10, 20]$ give
almost identical estimates of $\zeta(p)$ which are weakly
multifractal. For $Q=5$ the $\zeta(p)$ are shifted slightly toward
self similar scaling. The closeness of the conditioned results for
the range $Q=[5,20]$, and their clear separation from the raw
result, suggests that these are a reasonable approximate measure
of the scaling properties of the time series. This procedure can
be used to make quantitative comparisons between timeseries to
this precision. Given the caveats above however, we cannot use
this procedure to distinguish whether the time series is self
affine or weakly multifractal, but can distinguish strong
multifractality.

\begin{figure}[t]
\vspace*{2mm}
  \centering{\includegraphics[width=8.3cm]{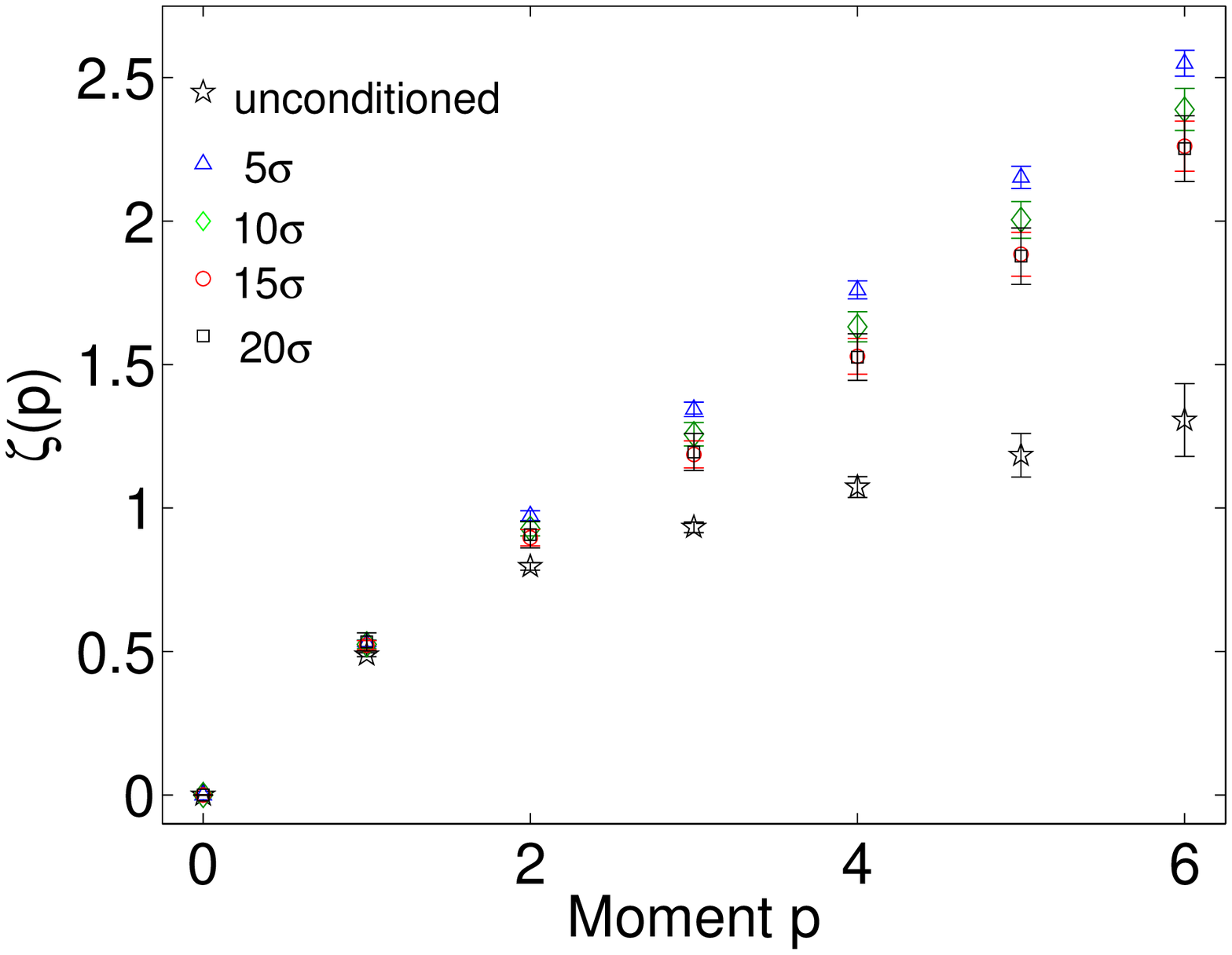}} 
  \caption{
    \label{fig:3}Scaling exponents $\zeta(p)$ versus $p$ for the
    AE index, shown in the same format as figure 2}
\end{figure}
\begin{figure}[t]
\vspace*{2mm}
  \centering{\includegraphics[width=8.3cm]{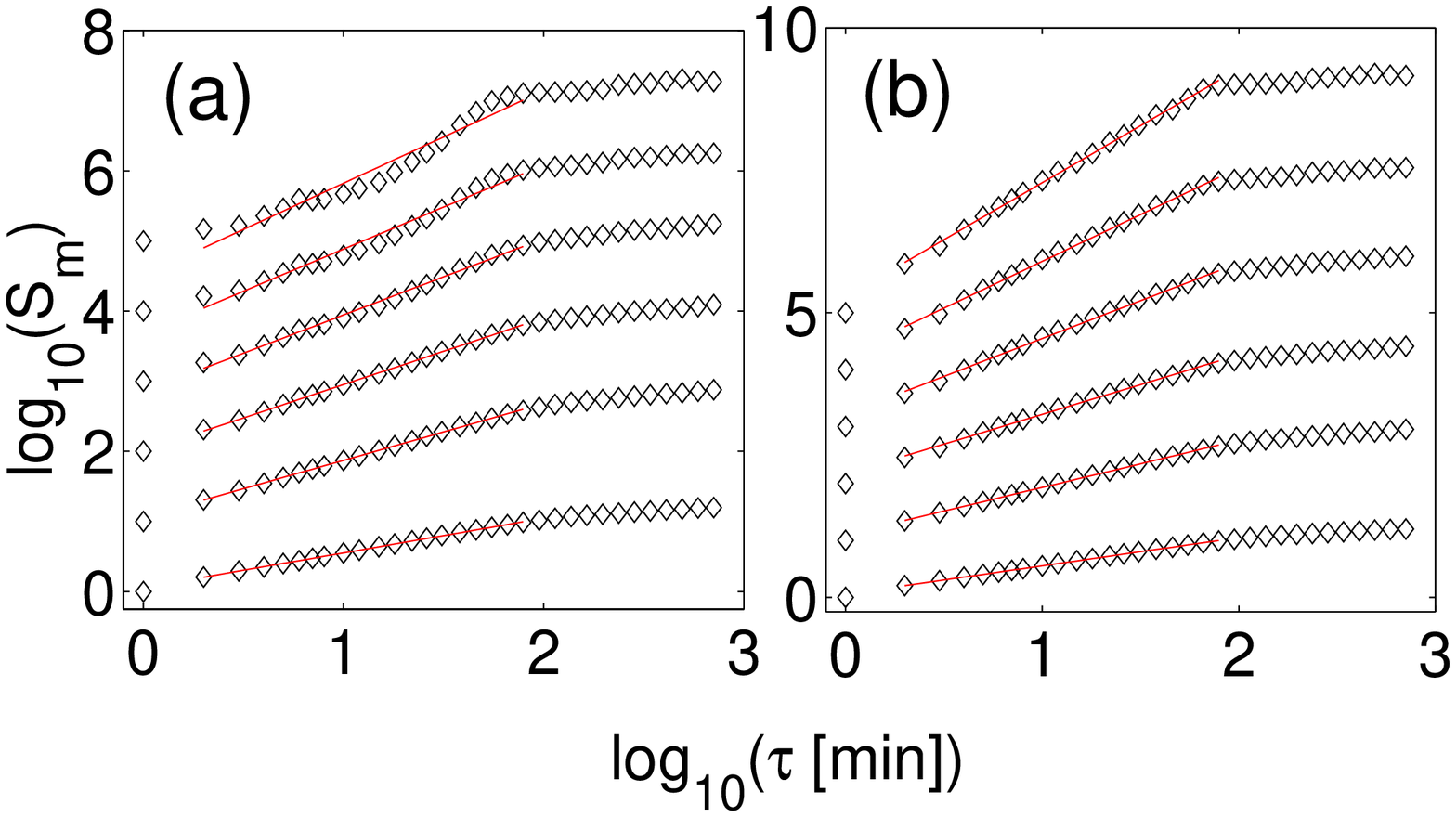}} 
  \caption{
    \label{fig:4}Structure functions of the AE index estimated for orders $p=[1,6]$
    by method
    (\ref{sfns}) (a) and by method (\ref{cond}) (b).
  }
\end{figure}
\section{Low Order Moments and Non Uniqueness: Comparison with a fractional Brownian surrogate.}

Equation (\ref{momexp}) expresses the PDF as an expansion in the
moments to all orders.  It follows that  distinct timeseries can
share the first few moments and therefore if scaling, may also
share the same Hurst exponent and corresponding exponent of the
power law power spectrum. Having estimated the scaling exponent of
the AE index as above we can construct a time series with the same
second moment from a fractional Brownian motion to illustrate
this.

The fractional Brownian walk was generated using the method
described in Appendix 3 of \cite{peters}. The algorithm takes a series of
Gaussian random numbers and approximates a finite correlation
time by weighting past values according to a power law function.
In our case 1024 Gaussian samples were used to create each
increment of fractional walk. The resulting time series is
comprised of $7.5 \times 10^5$ increments.

Figure 5 shows the two time series, (i) the interval of AE
analyzed above, and (ii) the fBm surrogate. The  standard
deviation versus $\tau$ for the two time series is shown in Figure
6. The power spectrum of AE (the raw, rather than the differenced
variable)( c.f. \cite{tsurutani,takalo93}), along with the
$\sigma(\tau)$ and the structure functions, show a characteristic
break on timescales of 1-2 hours. On times shorter than this, we
can obtain a scaling collapse of the PDF (see \cite{hnatgrl}, also
\cite{hnatjgr}. Fluctuations on these timescales share the same
second moment as the fBm. In Figure 7 we compare the PDF of these
fluctuations and we see that these are very distinct; fBm is
defined as having Gaussian increments (\cite{mandelbrot}) and this
is revealed by the PDF whereas the AE increments are non-Gaussian.

This is an illustration of the fact that the scaling in AE over
this region is not necessarily due to time correlation, the
``Joseph effect" for which Mandelbrot constructed fractional
Brownian motion as a model. Indeed AE has almost uncorrelated
differences at high frequencies, as indicated by its nearly
Brownian $f^{-2}$ power spectrum (\cite{tsurutani}). Rather the
scaling is synonymous with  the heavy tailed PDF (``Noah effect")
for which \cite{mandelbrot} earlier introduced a L\'{e}vy model in
economics.

\begin{figure}[t]
\vspace*{2mm}
  \centering{\includegraphics[width=8.3cm]{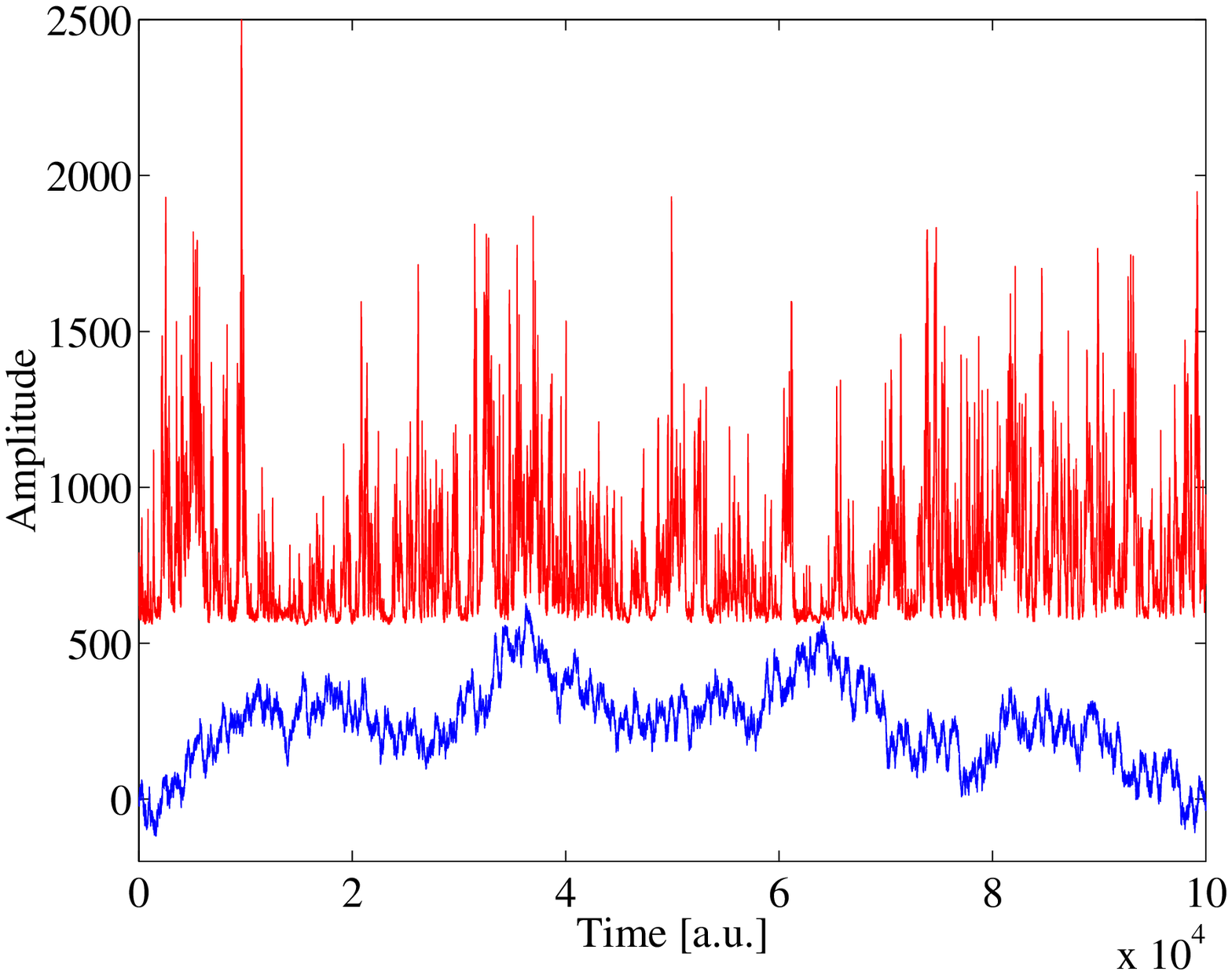}} 
  \caption{
    \label{fig:5}
    A $\sim 1.5$ year interval of AE data (upper trace) is shown alongside a surrogate
     fBm time series (lower trace) with the same second moment. The traces have been displaced for clarity.}
\end{figure}

\begin{figure}[t]
\vspace*{2mm}
  \centering{\includegraphics[width=8.3cm]{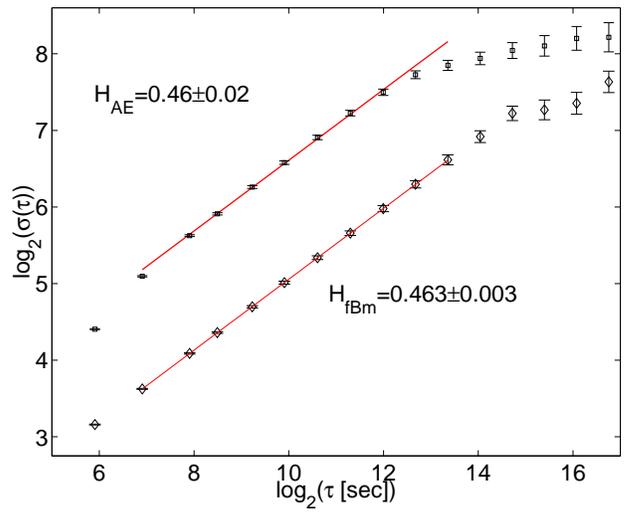}} 
  \caption{Standard deviation of the time series differenced on timescale $\tau$
  plotted versus $\tau$ for an interval of AE index data (see
  text) and an fBm time series constructed with the  same second moment. The traces have been displaced for clarity.
    \label{fig:6
}
    }
\end{figure}
\begin{figure}[t]
\vspace*{2mm}
  \centering{\includegraphics[width=8.3cm]{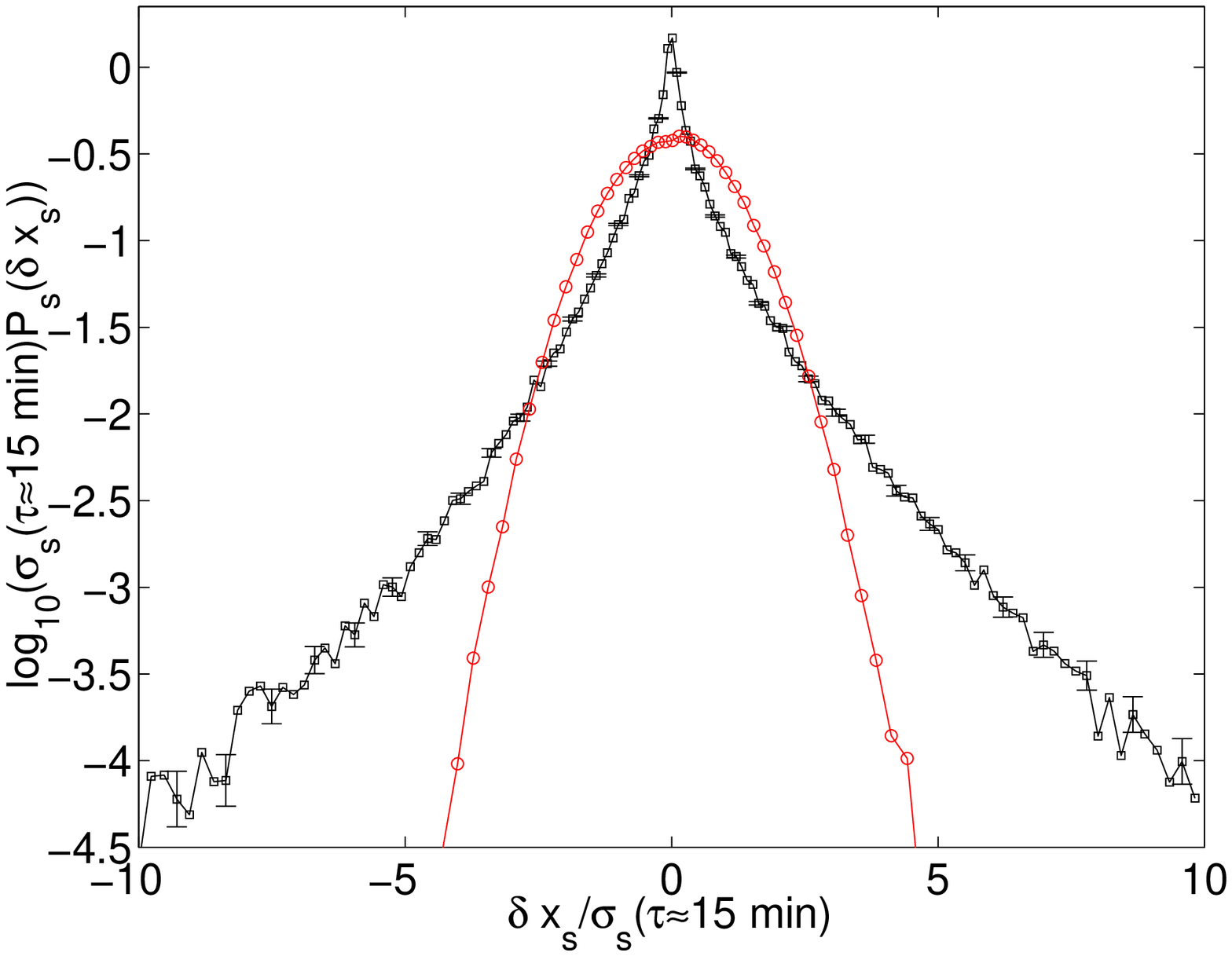}} 
  \caption{PDF of the time series of AE, differenced on timescales
  less than one hour ($\square$). The PDF of an fBm with the same second moment
  is shown for comparison ($\circ$).
    \label{fig:7}}\end{figure}

Finally, we plot in Figure 8 the $\zeta(p)$ versus $p$ obtained
from the structure function estimate (\ref{cond}) with $Q=10$ for
both time series. We see from the plot that both time series are
self affine and to within the uncertainty of the finite length
time series, both share values of $\zeta(p)$ for the lowest orders
in $p$. However the higher order structure functions reveal the
distinct scaling of the two time series.

\begin{figure}[t]
\vspace*{2mm}
  \centering{\includegraphics[width=8.3cm]{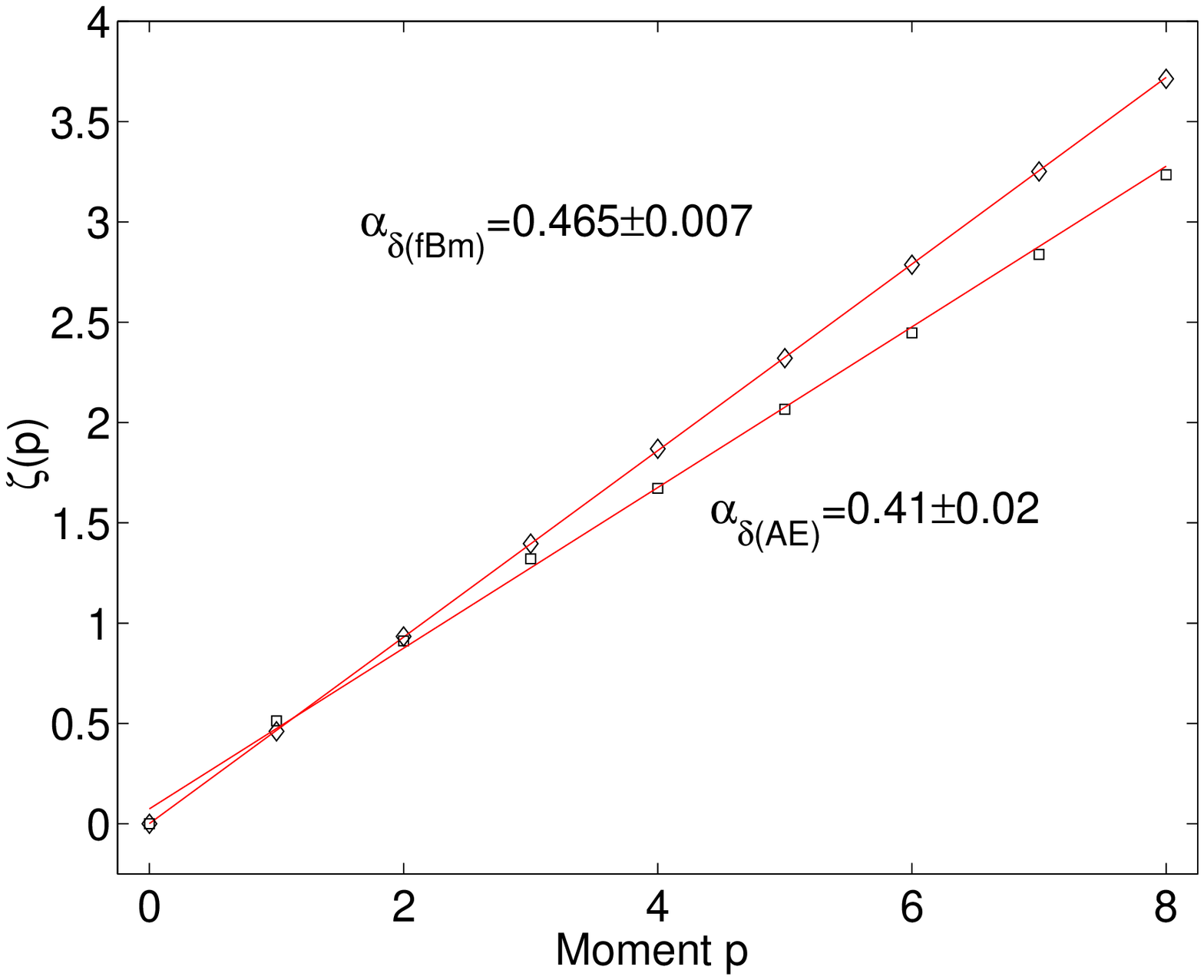}} 
  \caption{Structure functions obtained by conditioning at 10
  $\sigma$ for an interval of the AE index, and for a fBm
  constructed to share the same second moment.
    \label{fig:8}
    }
\end{figure}

\section{Fokker-Planck model} For completeness we now outline how the
exponent $\alpha$ of a self affine time series leads to the
functional form of $P(y)$ via a Fokker- Planck model of the
stochastic process $x(t)$. Here we will consider an approach where
scaling is achieved via transport coefficients that are functions
of the differenced variable $y(t)$. An alternative approach is via
fractional derivatives for the dependent (y) coordinate (see e.g.
\cite{schertzer,slez}). These are in principle equivalent (e.g.
\cite{yangeorge}).

We begin with a general form of the Fokker-Planck equation can be
written \cite{handbook}:
\begin{equation}
\frac{\partial{P}}{\partial{\tau}}= \nabla_{y} (A(y)P +
B(y)\nabla_{y}P), \label{f-p}
\end{equation}
where $P \equiv P(y,\tau)$ is a PDF for the differenced quantity
$y$ that varies with time $\tau$, $A(y)$ is the friction
coefficient and $B(y)$ is related to a diffusion coefficient which
we allow to vary with $y$. If we now impose the condition that
solutions of (\ref{f-p}) are invariant under the scaling given by
(\ref{rescale}), then it is found that both $A(y)$ and $B(y)$ must
have the form of power law dependence on $y$. Then as shown in
\cite{hnatpre}, (\ref{f-p}) takes the form:
\begin{equation}
\frac{\partial{P}}{\partial{\tau}}=\frac{\partial}{\partial{y}}\left[y\mid
y \mid^{-1/\alpha}\left(a_0P + b_0 y
\frac{\partial{P}}{\partial{y}}\right)\right], \label{mdl1}
\end{equation}
where $a_0$ and $b_0$ are constants, $\alpha$ is the scaling index
derived from the data and $P(y)$, $y$ are unscaled PDF and
fluctuations respectively, and where here we have explicitly
insisted that the diffusion coefficient $B(y)>0$. Importantly, in
a physical system the scaling behaviour (\ref{rescale}) is
expected to be strongly modified as $y\rightarrow 0$, that is, at
the peak of the PDF $P(y)$ since for a sufficiently small
difference between two measurements $x(t)$,
$y(t,\tau)=x(t+\tau)-x(t)$ will be dominated by the uncertainties
in those measurements.

Written in this form equation (\ref{mdl1}) immediately allows us
to identity $B(y) \propto y^2 \mid y \mid^{-1/\alpha}$ and
$A(y)\propto y\mid y \mid^{-1/\alpha}$. Solutions  to (\ref{mdl1})
exist which are functions of $y_s=y\tau^{-\alpha}$ only which
correspond to
 stationary solutions  with respect to $\tau$. We obtain
these by the change of variables ($P,y,\tau\rightarrow P_s,y_s$)
of (\ref{mdl1}):
\begin{equation}
\frac{b_0}{a_0}y_s\frac{dP_s}{dy_s}+P_s+\frac{\alpha}{a_0} \mid
y_s\mid^{\frac{1}{\alpha}}P_s = \frac{C \mid
y_s\mid^\frac{1}{\alpha}}{y_s}. \label{mdl1.5}
\end{equation}
This differential equation (\ref{mdl1.5}) can be solved
analytically with a general solution of the form:
\begin{eqnarray}
P_s(y_s)=\frac{a_0}{b_0}\frac{C}{|y_s|^{a_0/b_0}}
exp\left(-\frac{\alpha^2}{b_0}\mid
y_s\mid^{1/\alpha}\right)\nonumber
\\ \times \int_0^{y_s}
\frac{ \mid
y_s'\mid^{\frac{a_0}{b_0}}exp\left(\frac{\alpha^2}{b_0}\mid
y_s'\mid^{1/\alpha}\right)}{\mid y_s'\mid^{2-\frac{1}{\alpha}}}
d(y_s') + k_0H(y_s), \label{mdl2}
\end{eqnarray}
where $k_0$ is a constant and $H(y_s)$ is the homogeneous
solution:
\begin{equation}
H(\delta x_s)=\frac{1}{\mid y_s\mid^{a_0/b_0}}
exp\left(-\frac{\alpha^2}{b_0}\mid y_s \mid^{1/\alpha}\right).
\label{homsol}
\end{equation}
Power law scaling for arbitrary $y$ leads to singular behaviour of
this solution at $y\rightarrow 0$. We do not however expect this
to describe a physical system as $y\rightarrow 0$ as discussed
above. For (\ref{mdl2}) to describe a PDF we require that its
integral is finite. We can discuss this by considering the
behaviour close to the singularity:
\begin{equation}
\lim_{y_s\rightarrow 0}P(y_s) \simeq \frac{a_0}{b_0} \frac{C}{\mid
y_s\mid^\frac{a_0}{b_0}} \int^{y_s}_0 \frac{\mid
y_s'\mid^\frac{a_0}{b_0} dy_s'}{y_s'^{2-\frac{1}{\alpha}}}
+\frac{k_0}{\mid y_s\mid^\frac{a_0}{b_0}} =C+\frac{k_0}{\mid
y_s\mid^\frac{a_0}{b_0}}\label{limp}
\end{equation}
The integral of (\ref{limp}) is finite for $0\leq a_0/b_0<1$ and
$0 <\alpha \leq 1/2$ (a subdiffusive process) so that within this
range the integral of (\ref{mdl2}) will be finite also as
required. Outside of this range it can only be considered as an
asymptotic solution. However, we can consider the generalization
$y\rightarrow y+\epsilon$ in the above, where $\epsilon$ is a
constant of magnitude that is small compared to, say, the values
of $\sigma(\tau)$ for the physical system under study. This
eliminates the singular behaviour and corresponds (for $y$ small)
to the addition of low amplitude Gaussian noise as can be seen
from the form of the corresponding Langevin equation
(\ref{langevin}) below. Physically this corresponds to a simple
model for the statistical behaviour of the observational
uncertainties in the data which may dominate as the differenced
quantity $y \rightarrow 0$.

Expression (\ref{mdl2}) is then a family of solutions for the PDF
of self affine time series.  This provides a method to test for
self affinity that does not directly rely on determining the
scaling exponents to high order from the structure functions.
Having determined the exponent $\alpha$ from the scaling of a low
order moment (say, the standard deviation) one can then perform a
scaling collapse on the PDF; this should then also be described by
 the corresponding solution of (\ref{mdl2}) (see
 \cite{hnatpre,hnatjgr} for examples).

It is well known that a Fokker Planck equation is simply related
to a Langevin equation (see e.g. \cite{handbook}). A nonlinear
Langevin equation of the form
\begin{equation}
\frac{dy}{dt}=\beta(y)+\gamma(y)\xi(t) \label{langevin},
\end{equation}
where $\beta(y)$ is a $y$ -dependent force term and $\gamma(y)$ is
a $y$ -dependent noise strength, can be shown (\cite{hnatpre}) to
correspond to (\ref{f-p}) and in that sense to describe the time
series. In (\ref{langevin}) the random variable $\xi(t)$ is
assumed to be $\delta$-correlated, i.e.,
\begin{equation}
<\xi(t)\xi(t+\tau)>=\sigma^2 \delta(\tau). \label{dcorr}
\end{equation}
Consistency with equation (\ref{rg}) is achieved in the data
analysis by forming each time series $y(t,\tau)$ with
non-overlapping time intervals $\tau$.  Defining
$D_0=<\xi^2(t)>/2$ we then obtain:
\begin{equation}
\gamma(y)=\sqrt{\frac{b_0}{D_0}}y|y|^{-\frac{1}{2\alpha}},
\label{lngb}
\end{equation}
and
\begin{equation}
\beta(y)=\left[b_0(1-\frac{1}{2\alpha})-a_0\right]y|y|^{-\frac{1}{\alpha}}.
\label{lngg}
\end{equation}
With $\alpha=1/2$ and $a_0=0$ one recovers the Brownian random
walk with (\ref{f-p}) reduced to a diffusion equation with
constant diffusion coefficient.

Interestingly, \cite{beck} has independently proposed a nonlinear
Langevin equation where $\beta$ but not $\gamma$ varies with $y$.
This yields leptokurtic PDFs of the Tsallis functional form.

Finally the variable $\tau$ in (\ref{f-p}), and $t$ in
({\ref{langevin}) can be read in two ways: either as the
renormalization variable of the stochastic variable $y(t,\tau)$ or
the time variable of $x(t)$ since from (\ref{rg}) $\tau = 2^n
\Delta$ and with the choice $N=2^n$ we have $x(t)\equiv
y^n(t,\tau)$, $\tau \equiv t$ ($n,N$ large). Thus (\ref{langevin})
can be seen either as a prescription for  generating a self-
affine timeseries with scaling exponent $\alpha$, or as describing
the renormalization flow.

\conclusions\label{sec:end}

Empirical determination of the scaling properties and exponents of
time series $x(t)$ presents a formidable challenge in testing, and
developing, a theoretical understanding of turbulence and other
out-of-equilibrium phenomena. In this paper we have discussed the
special case of self affine time series by treating the
differenced variable $y(t,\tau)=x(t+\tau)-x(t)$ as increments of a
stochastic process (a generalized random walk). We have
highlighted two complementary approaches to the data.

The first of these is PDF rescaling; using a low order moment to
determine a scaling exponent and then verifying whether this
exponent collapses the PDFs of the differenced variable
$y(t,\tau)$ over the full range of $y$ accessible from the data.
As a corollary this collapsed PDF should also be well described by
the solution of a Fokker-Planck equation which has power law
transport coefficients.

The second of these is using structure functions to determine the
scaling properties of the higher order moments. In a finite length
time series the higher order structure functions can be  distorted
by isolated, extreme events which are not well represented
statistically. Using the example of a finite length L\'{e}vy
flight, we have demonstrated a method for conditioning the
time series that can in principle recover the underlying self
affine scaling.

Finally, to highlight how both these methods are complementary in
quantifying the scaling properties of the time series a fractional
Brownian walk was constructed to share the same second moment as
an interval of the differenced AE index time series. The two
timeseries were demonstrated to possess very different PDF of the
differenced variable, and distinct structure functions.

Both of these approaches could in principle be generalized to
multifractal time series (see e.g. \cite{schertzer}).

\begin{acknowledgements}

BH was supported by the PPARC. We thank John Greenhough, Mervyn
Freeman, and Murray Parkinson for stimulating discussions and the
UK Solar System Data Centre for the provision of geomagnetic index
datasets.

\end{acknowledgements}


\begin{thebibliography}{}
\bibitem[{\it Beck}(2001)]{beck}  Beck, C.,  Dynamical foundations of nonextensive statistical mechanics, {\em Phys. Rev. Lett., 87},
180601, 2001.

\bibitem[{\it Chapman and Watkins}(2001)]{chapman01} Chapman, S. C., and
N. W. Watkins, Avalanching and Self Organised Criticality: a
paradigm for magnetospheric dynamics?, {\em Space Sci. Rev., 95},
293--307, 2001.
\bibitem[\textit{Chechkin and Gonchar}, (2000)]{levy1} Chechkin, A. V., and V. Yu. Gonchar,
Self-Affinity of Ordinary Levy Motion, Spurious Multi-Affinity and
Pseudo-Gaussian Relations, \textit{Chaos, Solitons and Fractals},
11, 2379-2390, 2000

\bibitem[{\it Consolini et~al.}(1996)]{consolini96} Consolini, G., M. F.
Marcucci, M. Candidi, Multifractal structure of auroral electrojet
index data, {\em Phys. Rev. Lett., 76}, 4082--4085, 1996.
\bibitem[{\it Consolini and De Michelis}(1998)]{consolini98} Consolini, G.,
and P. De Michelis, Non-Gaussian distribution function of
$AE$-index fluctuations: Evidence for time intermittency, {\em
Geophys. Res. Lett., 25}, 4087--4090, 1998.

\bibitem[{\it Freeman et~al.}(2000)]{freeman} Freeman, M. P., N.~W. Watkins
and D.J. Riley, Evidence for a solar wind origin of the power law
burst lifetime distribution of the $AE$ indices,  {\em Geophys.
Res. Lett.,}{\em 27}, 1087--1090, 2000.
\bibitem[{\it Frisch}(1995)]{frisch} Frisch U.,{\em  Turbulence. The legacy of  A.N. Kolmogorov}, (Cambridge University Press, Cambridge, 1995).

\bibitem[{\it Gardiner}(1986)]{handbook}
Gardiner, C. W., \textit{Handbook of Stochastic Methods: For
Physics, Chemistry, and the Natural Sciences} (Springer Series in
Synergetics), Springer-Verlag, 1986.
\bibitem[{\it Hnat et~al.}(2002)]{hnatBgrl}Hnat, B., S. C. Chapman, G. Rowlands, N. W. Watkins, W. M. Farrell,
Finite size scaling in the solar wind magnetic field energy
density as seen by WIND, \textit{Geophys. Res. Lett}., 29, 86,
2002
\bibitem[{\it Hnat et~al.}(2003a)]{hnatgrl} Hnat, B., S. C.
Chapman, G. Rowlands, N. W. Watkins, M. P. Freeman, Scaling in
long term data sets of geomagnetic indices and solar wind
$\epsilon$ as seen by WIND spacecraft,  \textit{Geophys. Res.
Lett}.,{\em 30}, 2174, doi:10.1029/2003GL018209 2003a.
\bibitem[{\it Hnat et~al.}(2003b)]{hnatpre} Hnat, B., S. C. Chapman and G.
Rowlands, Intermittency, scaling, and the Fokker-Planck approach
to fluctuations of the solar wind bulk plasma parameters as seen
by the WIND spacecraft, {\em Phys. Rev. E} {\bf 67}, 056404,
2003b.
\bibitem[\textit{Hnat et al.} (2005)]{hnatjgr} Hnat, S. C. Chapman, G. Rowlands, Scaling and a Fokker-
Planck model for fluctuations in geomagnetic indices and
comparison with solar wind epsilon as seen by WIND and ACE,
\emph{J. Geophys. Res.}, in press, 2005

\bibitem[{\it Horbury and Balogh}(1997)]{horbury} Horbury
T.~S., and A. Balogh, Structure function measurements of the
intermittent MHD turbulent cascade, {\em Nonlinear Processes
Geophys., 4}, 185-199 1997.


\bibitem[{\it Kov\'{a}cs et~al.}(2001)]{kovacs} Kov\'{a}cs, P., V.~Carbone,
Z.~V\"{o}r\"{o}s, Wavelet-based filtering of intermittent events
from geomagnetic time series, {\em Planetary and Space Science,
49}, 1219-1231, 2001.


\bibitem[{\it Mandelbrot}(2002)]{mandelbrot} Mandelbrot, B.~B.,
{\em Gaussian Self-Affinity and Fractals: Globality, The Earth,
$1/f$ Noise and $R/S$}, (Springer-Verlag, Berlin, 2002).
\bibitem[{\it Mantegna and Stanley}(1995)]{stanley} Mantegna, R.~N., \& H.~E.
Stanley, Scaling Behavior in the Dynamics of an Economic Index,
{\em Nature, 376}, 46, 1995.

\bibitem[\it{Mangeney et al}, (2001)]{salem} Mangeney, A., C.~Salem, P.~L.~Veltri, and B.~Cecconi,  in {\em Multipoint measurements versus theory}, ESA report
SP-492, 492 (2001)

\bibitem[\textit{Nakao }(2000)]{levy2}
Nakao, H., Multiscaling Properties of Truncated Levy Flights,
\textit{Phys. Lett. A}, 266, 282-289, 2000
\bibitem[\it{Peters}, (1996)]{peters} Peters, E. E., \textit{Chaos and Order in the Capital Markets},
John Wiley and Sons, New York, New York, 1996.


\bibitem[{\it Schertzer et al}, (2001)]{schertzer} Schertzer, D., M. Larcheveque, J. Duan, V. V. Yanovsky
and S. Lovejoy, Fractional Fokker- Planck equation for nonlinear
stochastic differentisl equations driven by non- Gaussian L\'{e}vy
stable noises, {\em J. Math. Phys., 42}, 200--212, 2001.
\bibitem[\textit{Shlesinger et al}, (1995)]{slez}Shlesinger, M. F., G. M. Zaslavsky, U. Frisch (eds.),
 L\'{e}vy flights and related topics in physics: proc. int. workshop
 Nice, France, 27-30 June, 1994. Lecture Notes in Physics: 450. Springer-Verlag, Berlin, 1995.
\bibitem[\textit{Siegert and Friedrich}, (2004)]{sieg} Siegert, S.  and R. Friedrich, Modeling of L\'{e}vy
processes by data analysis, \textit{Phys. Rev. E}, 64, 041107,
2001.
\bibitem[{\it Sornette}(2000)]{sornette} Sornette, D., {\em Critical Phenomena
in Natural Sciences; Chaos, Fractals, Self- organization and
Disorder: Concepts and Tools}, Springer-Verlag, Berlin, 2000.
\bibitem[\textit{Sorriso-Valvo et al} (2001)]{valvo}
Sorriso-Valvo, L., V. Carbone, P. Giuliani, P. Veltri, R. Bruno,
V. Antoni and E. Martines, Intermittency in plasma turbulence,
\textit{Planet. Space Sci.} {\bf 49}, 1193--1200 2001.
\bibitem[{\it Stepanova et~al.}(2003)]{stepanova} Stepanova M.~V.,
E.~E.~Antonova, O.~Troshichev, Intermittency of magnetospheric
dynamics through non-Gaussian distribution function of PC-index
fluctuations, {\em Geophys. Res. Lett., 20} 30 (3), 1127 2003.
\bibitem[{\it Takalo et~al.}(1993)]{takalo93} Takalo, J., J. Timonen., and
H. Koskinen, Correlation dimension and affinity of $AE$ data and
bicolored noise, {\em Geophys. Res. Lett., 20}, 1527--1530, 1993.
\bibitem[{\it Takalo and Timonen}(1998)]{takalo98} Takalo J., and
J. Timonen, Comparison of the dynamics of the $AU$ and $PC$
indices, {\em Geophys. Res. Lett., 25}, 2101-2104, 1998.
\bibitem[{\it Tsurutani et~al.}(1990)]{tsurutani} Tsurutani, B. T.,{\it et~al.},
The nonlinear response of AE to the IMF $B_s$ driver: A spectral
break at $5$ hours, {\em Geophys. Res. Lett., 17}, 279--282, 1990.
\bibitem[\textit{Tu and Marsch,} (1995)]{tumarsh} Tu, C. -Y.and E.
Marsch, MHD Structures, waves and turbulence in the solar wind:
Observations and theories, \textit{Space Sci. Rev.} 73, 1, 1995.






\bibitem[{\it Uritsky and
Pudovkin}(1998)]{uritsky98} Uritsky V.~M., M.~I.~Pudovkin, Low
frequency 1/f-like fluctuations of the AE-index as a possible
manifestation of self-organized criticality in the magnetosphere,
{\em Annales Geophysicae 16 (12)}, 1580-1588, 1998.

\bibitem[{\it Uritsky et~al.}(2001)]{uritsky}
Uritsky,  V. M., A. J. Klimas and D. Vassiliadis, Comparative
study of dynamical critical scaling in the auroral electrojet
index versus solar wind fluctuations, {\em Geophys. Res. Lett.,
28}, 3809--3812, 2001.

\bibitem[{\it V\"{o}r\"{o}s et~al.}(1998)]{voros98} V\"{o}r\"{o}s, Z.,
P. Kov\'{a}cs, \'{A}. Juh\'{a}sz, A. K\"{o}rmendi and A. W. Green,
Scaling laws from geomagnetic time series, \textit{J. Geophys.
Res.,} {\em 25}, 2621-2624, 1998.
\bibitem[{\it V\"{o}r\"{o}s et~al.}(2002)]{voros02} V\"{o}r\"{o}s Z.,
D.~Jankovi\v{c}ov\'{a}, P.~Kov\'{a}cs, Scaling and singularity
characteristics of solar wind and magnetospheric fluctuations,
{\em Nonlinear Processes Geophys., 9 (2)}, 149-162, 2002.


\bibitem[{\it Watkins et~al.}(2001)]{watkins} Watkins, N.~W., M.~P. Freeman,
C.~S. Rhodes, G. Rowlands, Ambiguities in determination of
self-affinity in the $AE$-index time series, {\em Fractals, 9},
471-479, 2001.
\bibitem[{\it Weigel and Baker}(2003a)]{weigel03a}
Weigel, R. S.; Baker, D. N., Probability distribution invariance
of $1-$minute auroral-zone geomagnetic field fluctuations, {\em
Geophys. Res. Lett., 20} {\em 30}, No. 23, 2193,
doi:10.1029/2003GL018470 2003a.

\bibitem[{\it Wilson} (1979)]{wilson}Wilson, K. G. Problems in
physics with many scales of length, Scientific American, 241, 140,
1979.
\bibitem[{\it Yannacopoulos and
Rowlands}(1997)]{yangeorge}Yannacopoulos, A. N. and G. Rowlands,
Local transport coefficients for chaotic systems, {\em J. Phys. A
Math. Gen.} 30, 1503, (1997)

\bibitem[{\it Zaslavsky (1995)}]{zav} Zaslavsky, G. M., From Levy
flights to the Fractional Kinetic Equation for dynamical chaos, p
216, in Shlesinger, M. F., G. M. Zaslavsky, U. Frisch (eds.),
 L\'{e}vy flights and related topics in physics: proc. int. workshop
 Nice, France, 27-30 June, 1994. Lecture Notes in Physics: 450. Springer-Verlag, Berlin, 1995.

\end{thebibliography}
\end{document}